\documentclass[journal]{IEEEtran}
\ifCLASSINFOpdf
  \usepackage[pdftex]{graphicx}
\else
\fi
\usepackage{amsmath}
\usepackage{amssymb, amsfonts}

\usepackage{algorithmic}

\usepackage{hyperref}

\usepackage{amsmath,amssymb,amsfonts,bbm}
\usepackage{algorithmic}
\usepackage{graphicx}
\usepackage{textcomp}
\usepackage{multirow}
\usepackage{pifont,array,booktabs,colortbl,xcolor,makecell} % For check and cross marks
\usepackage{footnote}
\usepackage{printlen} 
\newcolumntype{C}[1]{>{\centering\arraybackslash}p{#1}}

\usepackage[style=ieee, maxbibnames=6, minbibnames=1]{biblatex}% including references instead of bibtex which is not working, urls still need to be deleted and maybe other formatting?

\def\BibTeX{{\rm B\kern-.05em{\sc i\kern-.025em b}\kern-.08em
    T\kern-.1667em\lower.7ex\hbox{E}\kern-.125emX}}

\begin{document}

% \title{\textbf{P}arallel \textbf{I}maging-Inspired \textbf{S}elf-\textbf{Co}nsistency: Improved Neural k-Space Representations for Dynamic MRI Reconstruction}
% \title{\textbf{P}arallel \textbf{I}maging-Inspired \textbf{S}elf-\textbf{Co}nsistency: \\ Self-Supervised k-Space Regularization for Improved Neural k-Space Representations 
% % for Dynamic MRI Reconstruction
% }
% \title{Parallel Imaging-Inspired Self-Consistency: k-Space Regularization of Neural Implicit k-Space Representations for Dynamic MRI}
\title{PISCO: Self-Supervised k-Space Regularization for Improved Neural Implicit k-Space Representations of Dynamic MRI}
% \title{Parallel Imaging-Inspired Self-Consistency: k-Space Regularization for Improved Neural Implicit k-Space Representations of Dynamic MRI}
% \title{Parallel Imaging-Inspired Self-Consistency: k-Space Regularization for Improved Neural Implicit k-Space Representations}

\author{Veronika Spieker, Hannah Eichhorn, Wenqi Huang, Jonathan K. Stelter, Tabita Catalan, \\
Rickmer F. Braren, Daniel Rueckert, Francisco Sahli Costabal, Kerstin Hammernik, \\
Dimitrios C. Karampinos, Claudia Prieto, Julia A. Schnabel
% \thanks{Manuscript received XXXXXX; accepted XXXXXX. Date of publication XXXXXX.}
\thanks{Veronika Spieker, Hannah Eichhorn and Julia A. Schnabel are with the Institute of Machine Learning for Biomedical Imaging, Helmholtz Munich, Neuherberg, Germany.}
\thanks{Veronika Spieker, Hannah Eichhorn, Wenqi Huang, Daniel Rueckert, Kerstin Hammernik, and Julia A. Schnabel are with the School of Computation, Information and Technology, Technical University of Munich, Germany  (e-mail: v.spieker@tum.de, hannah.eichhorn@tum.de, wenqi.huang@tum.de, khammernik@tum.de, julia.schnabel@tum.de).}
\thanks{Veronika Spieker, Tabita Catalan, Franciso Sahli Costabal and Claudia Prieto are with the Millenium Institute for Intelligent Healthcare Engineering, Santiago, Chile (e-mail: tcatalan@dim.uchile.cl, fsc@ing.puc.cl).}
\thanks{Jonathan K. Stelter, Rickmer Braren, Daniel Rueckert and Dimitrios Karampinos are with the School of Medicine and Health, Klinikum rechts der Isar, Technical University of Munich, Munich, Germany (e-mail: j.stelter@tum.de, rbraren@tum.de, daniel.rueckert@tum.de, dimitrios.karampinos@tum.de).}
\thanks{Daniel Rueckert is also with the Department of Computing, Imperial College London, London, United Kingdom.}
\thanks{Franciso Sahli Costabal and Claudia Prieto are with the School of Engineering, Pontificia Universidad Católica de Chile, Santiago, Chile.}
\thanks{Claudia Prieto and Julia A. Schnabel are also with School of Biomedical Imaging and Imaging Sciences, King’s College London, London, United Kingdom (e-mail: claudia.prieto@kcl.ac.uk).}
}

\maketitle

\begin{abstract}
Neural implicit k-space representations (NIK) have shown promising results for dynamic magnetic resonance imaging (MRI) at high temporal resolutions. Yet, reducing acquisition time, and thereby available training data, results in severe performance drops due to overfitting. 
To address this, we introduce a novel self-supervised k-space loss function $\mathcal{L}_\mathrm{PISCO}$, applicable for regularization of NIK-based reconstructions. The proposed loss function is based on the concept of parallel imaging-inspired self-consistency (PISCO), enforcing a consistent global k-space neighborhood relationship without requiring additional data.
Quantitative and qualitative evaluations on static and dynamic MR reconstructions show that integrating PISCO significantly improves NIK representations. Particularly for high acceleration factors (R$\geq$54), NIK with PISCO achieves superior spatio-temporal reconstruction quality compared to state-of-the-art methods. Furthermore, an extensive analysis of the loss assumptions and stability shows PISCO's potential as versatile self-supervised k-space loss function for further applications and architectures.
Code is available at: \url{https://github.com/compai-lab/2025-pisco-spieker} 
\end{abstract}


\section{Introduction}
\label{sec:introduction}
Magnetic resonance imaging suffers from long acquisition times, which can limit its spatial and temporal resolution. This particularly affects dynamic applications, where temporally resolved images are reconstructed by sorting the data into distinct motion states (MS), i.e. cardiac or respiratory motion states \cite{Zaitsev_2015,Spieker_2023}.
%%% Motion-resolved imaging (e.g. XD-GRASP)
Yet, the reconstruction of multiple MS reduces the available data per temporal MS, causing undersampling artefacts due to Nyquist theorem violations. Spatial reconstruction quality is commonly recovered by utilizing redundancies through regularization along the temporal dimension \cite{Feng_2016,Terpstra_2023}. Nonetheless, the limited number of motion states leads to motion blurring, caused by insufficient temporal resolution.

%%% Neural Implicit Representations 
Neural implicit representations have recently gained attention to learn continuous representations from discrete data \cite{Mildenhall_2020,Sitzmann_2020}, also for blurring-free dynamic MRI reconstructions \cite{Huang_2023,Spieker_2023_iconik,Catalan_2023,Kunz_2024,Shen_2022}. Using the acquired k-space acquisition trajectory and a temporal signal for the motion dimension, a multi-layer perceptron (MLP) is trained to predict the k-space or image signal corresponding to a given spatio-temporal input. Training the MLP exclusively in the k-space domain, as done in neural implicit k-space representations (NIK) \cite{Huang_2023,Spieker_2023_iconik}, allows for flexible, trajectory-independent training and avoids computationally expensive domain transforms, such as non-uniform Fast Fourier Transformations (NUFFT). 

%%% Limitation of NIK being self-supervised, requiring more data
%%% currently no regularization for NIK proposed
NIK is a self-supervised reconstruction method trained on a subject-specific basis (without the need for training data from other subjects). NIK's ability to work effectively with limited data is desirable for reducing individual acquisition times. In radial trajectories, which benefit from motion-averaging due to frequent sampling of the k-space center, such acceleration may lead to significant gaps towards the k-space periphery, where high-frequency details are represented. Consequently, without any additional regularization, NIK may be prone to overfitting and result in noisy reconstructions. General learning-based MRI reconstruction methods mitigate overfitting by incorporating regularization, typically applied in the image domain \cite{Ahmad_2020,Hammernik_2023,Jafari_2023,Huang_2021b}. Enforcing these image domain constraints directly would undermine the advantages of exclusive training in k-space, and translating these constraints into k-space is not trivial.

%%% general assumptions about global neighborhood relation in k-space can be made
% yet classically, the relation needs to be estimated on ACS
The classical parallel imaging concept Generalized Autocalibrating Partially Parallel Acquisitions (GRAPPA) \cite{Griswold_2002} builds on a potential spatial neighborhood relationship within the k-space itself. This relationship is initially estimated on a fully-sampled calibration set, allowing it to be applied to remaining undersampled k-space regions. This neighborhood relationship has already been leveraged in some learning-based reconstruction methods \cite{Ryu_2021,Spieker_2023_iconik}. Yet, like GRAPPA, they require explicit determination of the k-space neighborhood relationship. This requires a fully calibrated region for each MS, which is impractical for motion-resolved imaging.

We have recently introduced PISCO \cite{Spieker_2024_pisco}, a self-supervised k-space consistency condition that exploits the intrinsic global relationships within k-space, without requiring calibration data. However, PISCO was only tested as a loss function during NIK training on free-breathing data for dynamic MRI, without any validation of its assumptions and convergence behaviour. In this work, we conduct a comprehensive validation of the PISCO condition, investigating various design choices and demonstrating their convergence. 
% This includes a novel PISCO-based loss function, which we show to be more effective than the previously proposed version, i.e. regarding its convergence behaviour and improving NIK.
This includes a novel residual-based PISCO loss function with improved optimization properties and enhanced NIK performance compared to the previous version.
Compared to \cite{Spieker_2024_pisco}, we expand PISCO's evaluation by demonstrating its potential for MRI reconstruction in a broader setup, including a different dynamic MRI reconstruction problem (cardiac) as well as solving a distinct undersampled reconstruction problem (k-space optimization independent of NIK).
%%% Contributions %%% 
Overall, our contributions are three-fold:
\begin{itemize}
    % PISCO
    \item We present an improved concept of parallel imaging-inspired self-consistency (PISCO) \cite{Spieker_2024_pisco}, extended by a comprehensive analysis of key components such as kernel design, weight solving and consistency quantification.
    % PISCO-NIK loss
    \item We integrate PISCO in a novel self-supervised k-space loss function, validate its convergence behaviour compared to \cite{Spieker_2024_pisco} and assess its ability to enhance MRI reconstruction using NIK representations across three distinct in-vivo applications.
    \item The potential of PISCO is quantitatively and qualitatively demonstrated on static as well as multiple dynamic MRI applications using NIK, highlighting its ability to notably improve highly accelerated reconstructions.
\end{itemize}

%%%%%%%%%%%%%%%%%%%%%%%%%%%%%%%%%%%%%%%%%%%%%%%%%%%%%%%%%%%%%
\section{Background}
\label{sec:back}
This section introduces the parallel imaging-based k-space interpolation method GRAPPA \cite{Griswold_2002} and the k-space based dynamic reconstruction method NIK \cite{Huang_2023}. GRAPPA lays the foundation for our proposed self-supervised k-space condition PISCO, applicable as regularization for NIK.
\subsection{GRAPPA}
\label{sec:back_grappa}
\begin{figure}[!t]
\centerline{\includegraphics[width=1\columnwidth]{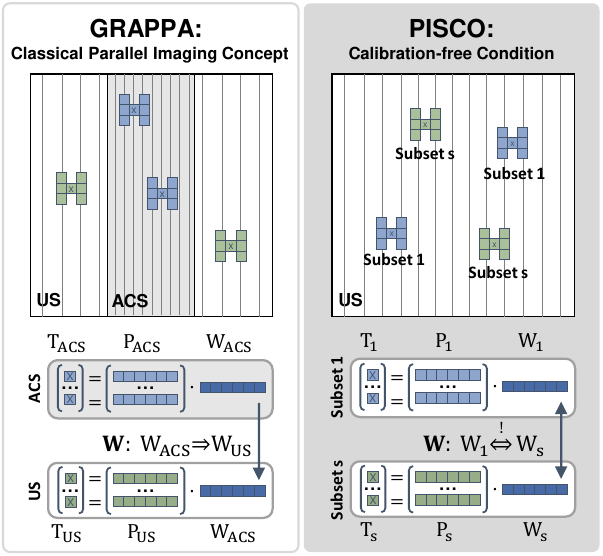}}
% \centerline{\includegraphics[width=\textwidth]{figures/test.pdf}}
\caption{From GRAPPA \cite{Griswold_2002} to PISCO (proposed): \textbf{GRAPPA} (left) calibrates a weight vector $\mathrm{W_{ACS}}$ on a fully sampled autocalibration k-space (ACS, gray area, Eq. \ref{eq:grappa_weightsolving}), where gray lines indicate sampled k-space points. This vector determines the global neighborhood relationship $\mathrm{W}$ and is applied to derive target points in the undersampled k-space (US, white area). For \textbf{PISCO} (right), multiple random subsets of targets and patches ($\mathrm{T}_s$ and $\mathrm{P}_s$) are sampled to solve for $\mathrm{W}_s$ (Eq. \ref{eq:PISCO_weightsolving}). The parallel-imaging inspired self-consistency (PISCO) condition states that for an ideal k-space, all these weight vectors should equally converge to one single global neighborhood relationship $\mathrm{W}$ (Eq. \ref{eq:PISCO_condition}).}
\label{fig:methods_g2p}
\end{figure}
%
% As explained in \cite{Spieker_2024_pisco}, 
An MR image ${x} \in \mathbbm{C}^{N_x \cdot N_y}$ is reconstructed by solving the inverse problem $y = \mathrm{A} x + n $, where $y=\{y_i\in \mathbbm{C}^{N_c}|i=1,...,N_x \cdot N_y\}$ is the k-space signal acquired with $N_c$ coils at spatial k-space coordinates $k=\{k_i \in \mathbbm{R}^{2}|i=1,...,N_x \cdot N_y\}$ (no time dimension for simplicity), $ \mathrm{A} = \mathcal{F} \mathbf{S} $ is the forward operator including coil sensitivity maps $\mathbf{S} \in \mathbbm{C}^{N_x \cdot N_y \cdot N_c}$ and Fourier transform $\mathcal{F}$ and $n$ is noise. In practice, only a part of the k-space $y$ is acquired (i.e. less $k_i$ sampled) to reduce acquisition time, which results in an undersampled dataset that violates the Nyquist theorem. To minimize subsequent undersampling artefacts, GRAPPA \cite{Griswold_2002} utilizes the multi-coil setup of MRI to estimate absent k-space values based on surrounding data points (see Fig. \ref{fig:methods_g2p} left). %\\
In detail, for a missing target location $k^\text{T}_{i} \in \mathbbm{R}^{2}$ a coordinate patch $k^\text{P}_{i} \in \mathbbm{R}^{N_n \cdot {2}}$ of $N_n$ neighboring coordinates is sampled. The missing target signal value $y^\text{T}_i \in \mathbbm{C}^{N_c}$ can then be estimated using a linear combination of the neighboring signal values $y^\text{P}_{i} \in \mathbbm{C}^{N_n \cdot N_c}$. % (Fig. \ref{fig:methods_g2p}). 
$N_m$ target/patch pairs can be stacked to create the linear equation system $ \mathrm{T} = \mathrm{P}\mathrm{W}$, where $\mathrm{T} = [y^\mathrm{T}_{1}, ..., y^\mathrm{T}_{N_m}] \in \mathbbm{C}^{[N_m \times N_c]}$, $\mathrm{P} = [y^\mathrm{P}_{1}, ..., y^\mathrm{P}_{N_m}] \in \mathbbm{C}^{[N_m\times N_n \cdot N_c]}$ and $\mathrm{W}\in \mathbbm{C}^{[N_n \cdot N_c\times N_c]}$ is the global weight matrix with a total of $N_w = N_n \cdot N_c \cdot N_c$ weights. Then, $\mathrm{W}$ can be determined on a fully sampled auto-calibration signal $y_{\mathrm{ACS}}$ by solving the following regularized least squares problem \cite{Griswold_2002}:
\begin{equation}
\label{eq:grappa_weightsolving}
  % \mathbf{W_{global}}: 
  \mathrm{W} \hat{=} \mathrm{W_{ACS}} = \arg \min_\mathrm{W}  \lVert \mathrm{P_{ACS} }\mathrm{W} - \mathrm{T_{ACS}} \rVert_2^2 +  \alpha \lVert \mathrm{W} \rVert_2^2.
  %\text{\quad s.t. } \mathrm{T}_{ACS}, \mathrm{P}_{ACS} \subseteq y_{ACS}.
\end{equation}
Here, $\mathrm{T_{ACS}}, \mathrm{P_{ACS}} \subseteq y_\mathrm{ACS}$, $\lVert \cdot \rVert_2^2$ applies the L2-norm element-wise and $\alpha$ weighs the Tikhonov regularization. Consequently, the missing k-space samples $y_{\mathrm{US}}$ are estimated by applying the determined weights $y^T_\mathrm{US} =\mathrm{W_{ACS}} \cdot y^P_\mathrm{US}$.

%%%%%%%%%%%%%%%%%%%%%%%%%%%%%%%%%%%%%%%%%%%%%%%%%%%%%%%%%%%%%
\begin{figure*}[t!]
\centerline{\includegraphics[width=\textwidth]{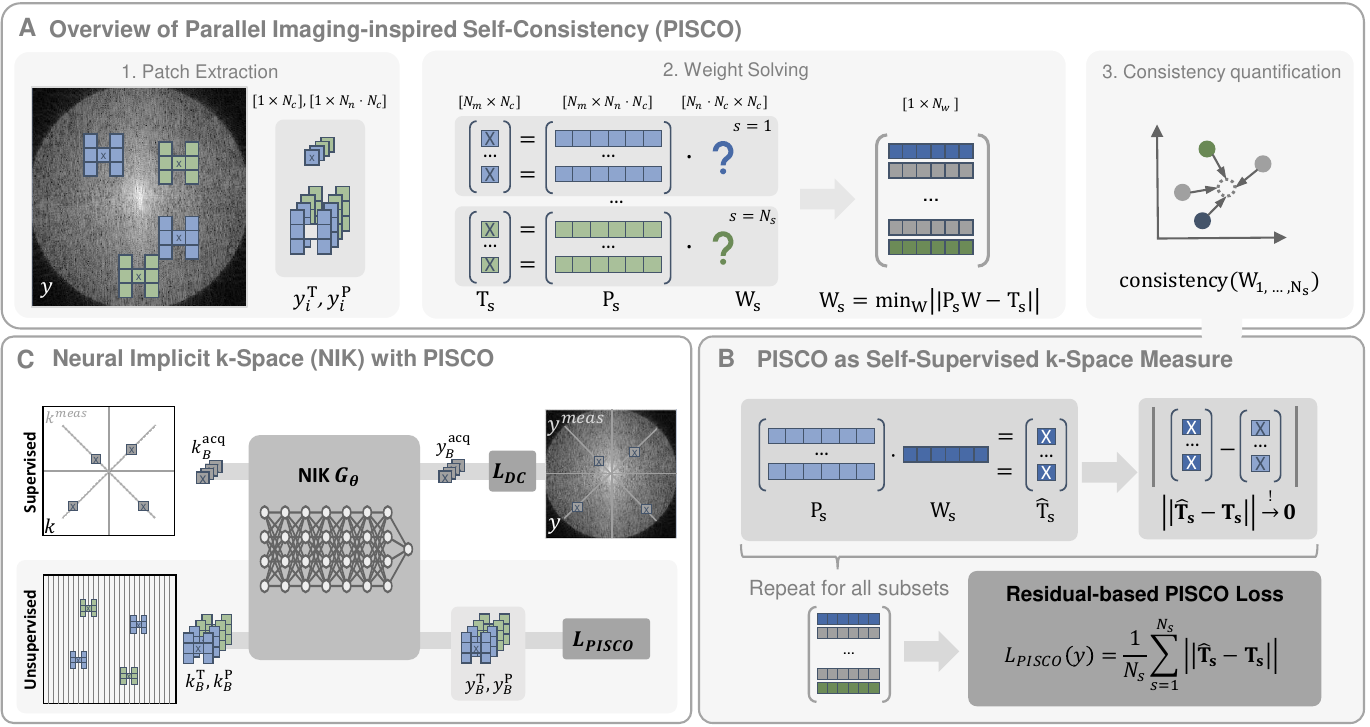}}
\caption{Method overview (A/B/C corresponding to Sec. \ref{sec:methods}A/B/C). For simplicity the coil dimension is not visualized, but indicated by the matrix sizes. (A) PISCO: 1. From a k-space $y$, multiple subsets (one color = one subset) including pairs of target $y_i^\text{T}$ and patches $y_i^\text{P}$ are sampled. The patches consist of neighboring values, whereas any sampling shape can be assumed (see Sec. \ref{sec:design_kernel}). 2. Each subset is reshaped to a linear equation system and used to solve for the neighborhood relationship vector $\mathrm{W}_s$ using Eq. \ref{eq:PISCO_weightsolving} (Sec. \ref{sec:design_weight}). 3. The PISCO self-consistency of $y$ can be quantified using all subset weight vectors, e.g. either computing the distance \cite{Spieker_2024_pisco} or the residual (proposed, visualized in B). (B) Residual-based PISCO as self-supervised k-space measure: For each subset $s$, $\mathrm{W}_s$ is used to estimate the targets $\Hat{\mathrm{T}}_s$ exclusively from their neighbors $\mathrm{P}_s$ and the residual to the actual targets $\mathrm{T}_s$ of $y$ is computed. The PISCO loss is determined as the weighted sum of all these residuals. (C) Integration of PISCO into NIK training: The "supervised" top represents the classical NIK-training \cite{Huang_2023}, where actually acquired coordinates are sampled, predicted and compared to the acquisition signal using $\mathcal{L}_{DC}$. With PISCO, the perceptual field of NIK can be extended during training, because any patches (independent of acquisition trajectory) can be sampled and used for the self-supervised $\mathcal{L}_\text{PISCO}$ computation. In the present study, patches are sampled from a Cartesian grid, with alternating undersampling in the x- and y-dimension (only in y-dimension visualized). 
% For simplicity, the coil dimension is omitted in the figure,  but needs to be considered during computation.
}
\label{fig:methods_overview}
\end{figure*}
%%%%%%%%%%%%%%%%%%%%%%%%%%%%%%%%%%%%%%%%%%%%%%%%%%%%%%%%%%%%%
\subsection{Neural Implicit k-Space Representation}
\label{sec:back_nik}
Reformulating the MRI reconstruction problem from Sec. \ref{sec:back_grappa} to the dynamic case with the temporal dimension $t$ results in k-space coordinates $k = \{k_i = [k_x, k_y, t] \in \mathbbm{R}^{3} |i=1,...,N_x \cdot N_y \cdot N_t\}$ and multiple temporal images $x=\{x_t|t=1,2,...,N_t\}$.
This problem can be solved using neural implicit k-space representations (NIK), which enable binning-free dynamic MR reconstructions at high temporal resolutions \cite{Huang_2008, Spieker_2023_iconik}. NIK consists of a model $G_\theta$ (usually a multi-layer perceptron) that maps input coordinates $k_i$ to multi-coil k-space signal values $y_i$ \cite{Spieker_2023_iconik}. During training, the model is fitted on a patient-specific basis on pairs of acquired k-space coordinates $k_i^\text{acq}$ and corresponding signal values $y_i^\text{acq}$: % by minimizing: 
\begin{equation}\label{eq:NIK_training}
    \theta^* = \arg \min_\theta \mathcal{L}_\text{DC}( G_\theta (k_i^\text{acq}), y_i^\text{acq})
\end{equation}
where $\mathcal{L}_\text{DC}$ is a data consistency loss and $\theta^*$ the optimized network parameters. At inference, any coordinate $k_i$ can be inputted into the fitted model $G_\theta^*$. The final dynamic reconstructions $x_t$ are obtained by sampling Cartesian k-space coordinates $\hat{k}$ 
%$k^{x,y=cart,t=t}$ 
at any time point $t$ and transforming them to image space, i.e., $x_t = \mathbf{S}^H \mathcal{F}^{-1} G_{\theta} (\hat{k})$.

%%%%%%%%%%%%%%%%%%%%%%%%%%%%%%%%%%%%%%%%%%%%%%%%%%%%%%%%%%%%%
\section{Methods}
\label{sec:methods}

\subsection{From GRAPPA to PISCO}
\label{sec:meth_g2p}
GRAPPA \cite{Griswold_2002} requires a fully sampled ACS for calibration of the global weight matrix $\mathrm{W}$. Yet, this is not always available, particularly in dynamic imaging. Hence, we reformulate the global spatial k-space relationship concept to a calibration-free condition based on a similar assumption as \cite{Spieker_2024_pisco}: 
If a weight set $\mathrm{W}_{\mathrm{ACS}}$ calibrated on $y_{\mathrm{ACS}} \in y $ models the global linear relationship $\mathrm{W}$ for the whole k-space $y$, then a weight set $\mathrm{W}_\mathrm{s}$ derived from a random subset $y_\mathrm{s} \in y $ should equally result in a global linear relationship within an ideal k-space:
\begin{equation}\label{eq:PISCO_weightsolving}
   \mathrm{W } \overset{!}{=} 
   \mathrm{ W}_s = \arg \min_{\mathrm{W}} \lVert \mathrm{P}_s \mathrm{W} - \mathrm{T}_s \rVert^2_2 +  \alpha \lVert \mathrm{W} \rVert^2_2 
   % \text{\quad s.t. } \mathrm{T}_s, \mathrm{P}_s \subseteq y_s . %, \mathrm{P}_s = \mathcal{P}(\mathrm{T}_s) %\in y_s
\end{equation}
where $ \mathrm{T}_s, \mathrm{P}_s \subseteq y_s$. Considering one global linear k-space relationship, multiple weight sets $\mathrm{W}_1, ..., \mathrm{W}_{N_s}$ derived from various random subsets $\{y_\mathrm{s} \in y| s= 1,...,N_s\}$ (visualized in green and blue in Fig. \ref{fig:methods_g2p}) are expected to converge to the same solution. 
% Thus, \textbf{P}arallel \textbf{I}maging-inspired \textbf{S}elf-\textbf{Co}nsistency (\textbf{PISCO}) is given if 
Thus, without access to fully-sampled calibration data, \textbf{P}arallel \textbf{I}maging-inspired \textbf{S}elf-\textbf{Co}nsistency (PISCO) of an undersampled k-space can be enforced as follows:
\begin{equation}\label{eq:PISCO_condition}
% \textbf{PISCO: } 
\mathrm{W}_1 \overset{!}{=} ... \overset{!}{=} \mathrm{W}_{N_s} ( \overset{!}{=} \mathrm{W} ) . 
\end{equation}
% \subsection{PISCO as k-Space Consistency Measure}

\subsection{PISCO as Self-Supervised k-Space Regularization}
\label{sec:meth_piscoreg}
Next to checking the consistency of a given k-space, PISCO can also serve as a self-supervised regularization objective function by quantifying the PISCO condition within a loss function $\mathcal{L}_\text{PISCO}$. Therefore, a proper computation of the consistency of all the solved subset weight vectors $\mathrm{W}_s$ is required (Fig. \ref{fig:methods_overview}A.3). In \cite{Spieker_2024_pisco}, the PISCO of a k-space $y$ is quantified by summing the complex distance $\mathcal{L}_{\mathbbm{C}}^1(\Delta) = ||\mathit{Re}(\Delta)||_1 + ||\mathit{Im}(\Delta)||_1$ between all estimated weight vectors:
\begin{equation}
    \mathcal{L}_\text{PISCO-dist} = \frac{1}{N_s^2} \sum_{i=1}^{N_s} \sum_{j=1, j\neq i}^{N_s} 
    \mathcal{L}_{\mathbbm{C}}^1(\mathrm{W}_{i} - \mathrm{W}_{j}).
\end{equation}
To avoid the risk of convergence to an unfeasible global solution and improve the condition's stability, we alternatively propose to enforce PISCO by measuring the fit of a determined weight set $\mathrm{W}_\mathrm{s}$ as follows (Fig. \ref{fig:methods_overview}B):
Consider a target and patch matrix $ \mathrm{T}_s, \mathrm{P}_s \subseteq y_s$ with more target/patch pairs $N_m$ than unknown weights $N_w$, i.e., an overdetermined linear equation system (LES) when solving for $\mathrm{W}_\mathrm{s}$ with Eq. \ref{eq:PISCO_weightsolving}. For an ideal k-space, the determined $\mathrm{W}_\mathrm{s}$ should consistently model the global neighborhood relationship over the whole k-space. Thus, the neighborhood-derived k-space $\hat{\mathrm{T}}_s = \mathrm{P}_s \mathrm{W}_s$ should approximate the original targets $\mathrm{T}_s$:
% For an ideal k-space the resulting residual should approximate zero, since a consistent linear relationship between all pairs is expected. The novel residual-based PISCO condition can be formulated as: 
\begin{equation}\label{eq:PISCO_residual_condition}
% \textbf{PISCO: }  \lVert \mathrm{P}_s \mathrm{W}_s - \mathrm{T}_s \rVert \xrightarrow{!} 0.
% \textbf{PISCO: }  \lVert \mathrm{P}_s \mathrm{W}_s - \mathrm{T}_s \rVert  = \lVert \hat{\mathrm{T}}_s - \mathrm{T}_s \rVert \xrightarrow{!} 0.
\textbf{PISCO: }  \lVert \hat{\mathrm{T}}_s - \mathrm{T}_s \rVert_p \xrightarrow{!} 0.
\end{equation}
%% Global integration into MRI recon? Similar to LORAKS/Espirit
The computation of the residual with any $p$-norm enables the quantification of PISCO for a given k-space, e.g the larger the residual the lower the self-consistency and vice versa. 
% Following, this condition can be integrated into the objective of a k-space-based reconstruction problem as regularization term: 
This condition can be rephrased as regularization loss $\mathcal{L}_\text{PISCO}$ for a given k-space $y$ (Eq. \ref{eq:PISCO_loss}) and integrated into the reconstruction objective, combining the NIK and PISCO loss terms (Eq. \ref{eq:PISCO_recon}): 
\begin{equation}\label{eq:PISCO_loss}
    \mathcal{ L}_\text{PISCO}(y) = \frac{1}{N_s} \sum_{s=1}^{N_s}  (\lVert \underbrace{\mathcal{K}_{\mathrm{P}}(y_{s}) \mathrm{W}_{s}}_{\hat{\mathrm{T}}_s} 
    - \underbrace{\mathcal{K}_{\mathrm{T}}(y_{s})}_{\mathrm{T}_s} \rVert_2)
\end{equation}
\begin{equation}\label{eq:PISCO_recon}
    \mathcal{ L}_\text{RECON}(y) = \mathcal{ L}_\text{DC}(y) + \lambda \cdot \mathcal{ L}_\text{PISCO}(y)
\end{equation}
where $y_s \in y$, $\mathcal{K}_{\mathrm{P}}$ and $\mathcal{K}_{\mathrm{T}}$ represent kernels extracting the patches and targets from $y_{s}$ according to Sec. \ref{sec:back_grappa}, $\mathrm{W}_s$ is the estimated neighborhood relationship (Eq. \ref{eq:PISCO_weightsolving}), $\mathcal{ L}_{DC}$ can be any loss enforcing data consistency and $\lambda$ is a weighting factor. In the following, PISCO will refer to the proposed residual-based version and PISCO-dist to the distance-based version.

% \subsection{Learning with PISCO}
\subsection{Neural Implicit k-Space Representations with PISCO}
% benefit of training NIK with PISCO
NIK are fitted to the acquired data (Sec. \ref{sec:back_nik}) only,
% using a data consistency loss $L_{DC}$ (see Fig. \ref{fig:methods_overview}, 
which limits the training strategy to coordinates from the acquisition's trajectory. Including PISCO allows for self-supervision of more coordinates independent of the acquired data within the training strategy, thereby enhancing the receptive field during the training procedure. This is particularly beneficial for large gaps within k-space, e.g. in radial trajectories.

When training NIK with PISCO, a batch of acquired coordinates $k_B^\text{acq}$ as well as further coordinate batches of target $\Tilde{k}_B^\text{T}$ and respective patch neighbors $\Tilde{k}_B^\text{P}$ are inputted to the NIK. Due to NIK's continuous sampling nature, any target and kernel for neighbor extraction can be chosen, e.g. as shown in Fig. \ref{fig:methods_overview} in a Cartesian manner. As in standard NIK \cite{Huang_2023}, $y_B^\text{acq} = G_\theta (k_B^\text{acq})$ is used for $\mathcal{L}_\text{DC}$ computation. Simultaneously, $\Tilde{y}_B^\text{T}$ and $\Tilde{y}_B^\text{P}$ are sorted into $N_s$ subsets with $N_m = f_{od} \cdot N_w$ patch pairs each, where $f_{od} > 1$ is a factor to ensure an overdetermined LES. The resulting subsets of patch pairs are then processed to compute $\mathcal{L}_\text{PISCO}$ (Eq. \ref{eq:PISCO_loss}) and the overall objective (Eq. \ref{eq:PISCO_recon}).
% Overall, the model is trained using the summed loss $\mathcal{L}_{NIK} = \mathcal{L}_{DC} + \lambda \mathcal{ L}_\text{PISCO}$, where $\lambda>0$ weighs the PISCO regularization strength. 
To ensure reasonable weight estimates for the PISCO computation, pre-conditioning of the model is recommended, i.e., by setting $\lambda=0$ up to a specified epoch $E_\text{pre}$ during training.

% Pre-training / pre-conditioning
% Outlier robustness with alpha in weight solving?
% overdetermination factor $N_m = f_{od} \cdot N_w$, with $f_{od}>1$ 

% \begin{equation}\label{eq:PISCO_NIK_training}
%     \mathcal{L}_{PISCO} = \mathcal{L}_{DC}(y_B, y_B^{acq} + \lambda \mathcal{L}_{PISCO}{}
% \end{equation}

%
%
%%%%%%%%%%%%%%%%%%%%%%%%%%%%%%%%%%%%%%%%%%%%%%%%%%%%%%%%%%%%%
\renewcommand{\arraystretch}{1.15} % Adjust row spacing
\begin{table*}[!t]
\caption{Summary of Acquisition Parameters and Evaluation Settings}
\centering
\begin{tabular}{cl|C{41mm}|C{41mm}|C{41mm}}
                                    &   & \textbf{Upper leg}  & \textbf{Cardiac cine} \cite{ElRewaidy_2020} & \textbf{Abdominal}   \\
\Xhline{1pt} %\hline \hline
\multirow{7}{*}{\rotatebox[origin=c]{90}{Acquisition}} 
                                        & Pulse Sequence       & spoiled GRE    & bSSFp cine % (25MS)      
                                        & spoiled GRE   \\
                                      \cline{2-5}
                                        & Trajectory       & pseudo golden angle stack-of-stars    & radial (binned) % (25MS)      
                                        & pseudo golden angle stack-of-stars   \\
                                      \cline{2-5}
                                      & Resolution  &       
                                      1.5 × 1.5 × 3 mm$^3$ &       
                                      % 380 × 380 mm$^2$ /  
                                      1.8 × 1.8 mm$^2$    &    
                                      1.5 × 1.5 × 3 mm$^3$  \\
                                      \cline{2-5}
                                      & Number of coils $N_c$      & 27      & 15-18     & 26    \\
                                      \cline{2-5}
                                      & \begin{tabular}[c]{@{}l@{}} Number of k-space points \\ $[N_{spokes}$ x $N_{FE}]$ per slice \end{tabular} & 704 × 448 &  
                                      \begin{tabular}[c]{@{}c@{}}4900 × 414 (total) \\ 196 × 414 (per MS) \end{tabular}  &     1341 × 536                                                                                                               \\
                                    \cline{2-5}
                                      & Acceleration Factor per MS         & {R1} & R1 (per MS) & $\sim$R0.6/R2.4/R30 (1/4/50MS)   \\
\Xhline{1pt} %\hline \hline
\multirow{4}{*}{\rotatebox[origin=c]{90}{Evaluation}}  
                                    & Reconstructed MS  & 1  & 25 & 4/50$^1$   \\
                                       \cline{2-5}
                                    & Reconstruction matrix  & 268 × 268   & 208 × 208$^2$, cropped to 104 × 104 %$\xrightarrow[]{}$ 104 × 104   
                                    & 268 × 268   \\
                                    \cline{2-5}
                                      & Reference reconstruction                            & XD-GRASP1 for R1 ($\lambda_{TV}$=0.1)    & XD-GRASP25 for R1 ($\lambda_{TV}$=0.01) & -                           \\
                                    \cline{2-5}

                                      & Tested accelerations$^3$          & R5, R10, R20  ($\lambda_{TV}$=0.1)   & R15, R26, R52  ($\lambda_{TV}$=0.3)     & $\sim$R30    ($\lambda_{TV}$=0.1)                        \\
\Xhline{1pt} %\hline \hline
\end{tabular} \\
\vspace{0.5mm}
\raggedleft \footnotesize{$^1$4MS as in \cite{Feng_2016}, 50MS for high temporal resolution $^2$after zerofilling removal \cite{Shimron_2022}} $^3$With $\lambda_{TV}$ for accelerated XD-GRASP reconstructions
\label{table:imaging_parameters}
\end{table*}
%
%%%%%%%%%%%%%%%%%%%%%%%%%%%%%%%%%%%%%%%%%%%%%%%%%%%%%%%%%%%%%
\section{Experimental Setup}
\label{sec:experimentalsetup}
\subsection{Data}
\label{sec:data}
Experiments are conducted on three different MR datasets of varying complexity, with detailed acquisition parameters for all datasets listed in Table \ref{table:imaging_parameters}. In-house data was acquired at 3T (Ingenia Elition X, Philips Healthcare, Best, The Netherlands) after local ethics committee approval. Coil sensitivities are estimated with ESPIRiT \cite{Uecker_2014}. The datasets evaluated are: \\
\textbf{Upper leg (quasi-static/no motion)} 
A quasi-static radial stack-of-stars in-house acquisition of the upper leg to validate PISCO's potential independent of the time dimension. 
\\
\textbf{Cardiac cine (cardiac motion)} 
A public cardiac cine dataset \cite{ElRewaidy_2020} with binned fully-sampled data as reference (retrospective ECG-triggering). For 30 subjects in total, zero-padding is removed before processing to avoid implicit data crimes \cite{Shimron_2022} and retrospective undersampling is conducted to match a uniform distribution per MS \cite{ElRewaidy_2021}.
\\
% \paragraph{Abdominal Simulation (respiratory motion)} 
% To obtain a reference for abdominal motion-resolved imaging, we create a dynamic free-breathing simulation of the abdomen with the XCAT phantom \cite{Segars_2010}. First, water/fat/susceptibility \cite{Collins_2002,Maril_2005} maps are generated for 100 time points $t_{MS}$ within one breathing cycle. Using a water-fat model \cite{Yu_2008} (echo time $T_e$=1.4ms), complex images $\textbf{x}_{t}$ are generated. With six simulated coil sensitivity maps and the NUFFT, $\textbf{x}_{t}$ is transformed to radial motion-free k-space data $Y_{mf}$ for R=1,2,3. Based on a respiratory navigator obtained from the lung-liver edge, $Y_{mf}$ is finally combined into one motion-affected k-space $Y_{ma}$.
\textbf{Abdominal (respiratory motion)}
A free-breathing radial stack-of-stars in-house acquisition of the abdomen and, for comparison, a gated acquisition of the same subject.

\subsection{PISCO Design Choices}
\label{sec:design_choices}
Within PISCO, multiple design choices can be made to ensure and improve PISCO convergence behaviour. In the following, we explain major design choices for each module in the PISCO computation pipeline: (1) patch extraction (Fig. \ref{fig:methods_overview}A.1), (2) weight solving (Fig. \ref{fig:methods_overview}A.2) and (3) PISCO consistency quantification (Fig. \ref{fig:methods_overview}A.3). Therefore, we experimentally validate the effect of different design choices independently of NIK on a sample frame for a \textbf{cardiac cine} subject by simulating ideal k-space data, i.e., generating signal values for all required k-space coordinates using \textit{torchkbnufft} \cite{Muckley_2020}.
% We experimentally validate the effect of different design choices independent of NIK. Therefore, we simulate an ideal k-space for an exemplary frame of a \textbf{cardiac cine} subject by obtaining the signal values for all required k-space coordinates using \textit{torchkbnufft} \cite{Muckley_2020} for the following experiments.
% To ensure convergence of PISCO and determine suitable design choices, we investigate the effect of different design choices for all three steps in the PISCO computation pipeline:  For evaluation purposes, we simulate the signal intensity values at the required k-space coordinates using NUFFT on a phantom MR image.

\subsubsection{Kernel Design}
\label{sec:design_kernel}
% \paragraph{Spatial Kernel Design}
% Cartesian vs. radial vs. equidistant
The combination of PISCO and NIK allows for an arbitrary selection of target points $\mathrm{T}$ as well as any kernel design $\mathcal{K}_{\mathrm{P}}$ to extract the neighboring patches $\mathrm{P}$, since it is not necessary that the kernel points are actually sampled within the training dataset. The target points $\mathrm{T}$ are selected to be on a Cartesian grid to focus the model's attention to sparsely sampled regions. The neighboring points $\mathrm{P}$ could be chosen arbitrarily, yet the assumed global consistency needs to be ensured.
% To increase the receptive field of the model during training, we sample target points in a Cartesian manner, thereby filling large radial gaps.
% Still, the assumed global consistency needs to be ensured. 
Therefore, we investigated three spatial kernel geometries extracting $\mathrm{P}$ around $ \mathrm{T}$: (a) Cartesian kernels, as originally used in GRAPPA \cite{Griswold_2002}, (b) radial kernels, as proposed in \cite{Seiberlich_2011} for radial trajectories and (c) equidistant radial kernels, which mitigate unequal kernel spacing at different radii of the radial sampling. We test kernels of shape $[3\times2]$ \cite{Griswold_2002} with an empirically determined neighbor distance of $\delta= 2 \cdot N_{FE}^{-1}$ in Cartesian or radial coordinates for the equidistant and radial kernel, respectively. To avoid temporal blurring due to merging of multiple time points \cite{Breuer_2005}, all patches within one subset are sampled from one time point. The global consistency assumption is then validated by sampling multiple random subsets $y_s$ from the artificially generated k-space with the different kernel geometries. %The intensity values are artificially generated using NUFFT for the required trajectory points. 
Then, the resulting subset weight vectors $\mathrm{W}_s$ can be compared and - since we assume an ideal k-space - all weight vectors should resemble each other. 
%% Dynamic MRI usually with radial trajectory
%% classical GRAPPA initially developed with Cartesian, but has been extended to radial kernels. Radial kernels pose the problem

\subsubsection{Weight Solving}
\label{sec:design_weight}
The computation of the PISCO residual requires the LES to be overdetermined by a factor $f_{od}=N_m/N_w$ larger than 1. To ensure robustness of the solution albeit the overdetermination, i.e. to outliers, regularization of the weight vector magnitude is included and weighted by a factor $\alpha$, as stated in Eq. \ref{eq:PISCO_weightsolving}. Empirical evaluation yielded feasible values of $\alpha$=1e-4 and $f_{od}$=1.1 for all datasets.

Another challenge in weight solving arises from the high dynamic range of k-space magnitudes (varying from center to the periphery k-space). Mixing patches with these different magnitudes results in a poorly scaled LES, posing an instability risk. Thus, randomly sampled patches are sorted according to their k-space center distance before being divided into subsets, thereby minimizing the magnitude variance within the LES for each individual subset. Additionally, k-space points within a small radius $r$ around $k_x/k_y$=0 are removed to avoid inclusion of individual high magnitudes (in our case $r = 10 \cdot N_{FE}^{-1}$, requires adjustment if kernel size and distance are increased). Using the k-space obtained with \textit{torchkbnufft}, we validate the effect of patch sorting on the weight estimates. 
% requires an overdetermined LES not only to enable computation of a plausible residual for PISCO, but also ensure robustness of the weight vector solution. Next to $f_{od}=N_m/N_w$ determining the overdetermination rate, regularizing the magnitude of the weight vector solution with the weighting factor $\alpha$ ensurest vector solution within  Eq. \ref{eq:PISCO_weightsolving}.  
% overdetermination
% tikhonov regularization parameter (plot different ones)
% sorting of k-space from center to outside
% mini-batching for parameter reduction 

\subsubsection{PISCO Consistency Measure}
\label{sec:design_measure}
As presented in Sec. \ref{sec:meth_piscoreg}, the self-consistency of a k-space can be measured in several ways. Yet, for ideal convergence behaviour during training the loss must be monotonically decreasing when approximating an ideal k-space. To test the convergence behaviour, we create "non-ideal" k-space data by (1) adding noise directly in k-space and (2) adding noise in image-space before applying the Fourier transform. We add zero-centered complex Gaussian noise at different standard variations $\sigma$, thereby determining the noise/corruption level. Then, both PISCO measures, distance-based \cite{Spieker_2024_pisco} and residual-based (proposed), are evaluated dependent on the noise level.
% Inspired by ESPiRiT \cite{Uecker_2014} and LORAKS \cite{Haldar_2014}, we reformulate PISCO into a loss that computes a residual of the predicted target points $T$ and the neighborhood-derived target points $T_{est}$.
% For loss computation we propose a more stable variant, that minimizes the residual of the solved weight vectors
% \subsection{$\mathcal{L}_{PISCO}$ Design Choices}
% \label{sec:data}
% \subsubsection{Integration into NIK}
%% it this too much? just describe in training and inference?

\subsection{Validation of PISCO Convergence}
To ensure convergence of PISCO independent of NIK, we provide a proof-of-concept by solving a simplified version of the PISCO reconstruction problem in Eq. \ref{eq:PISCO_recon}. Instead of $\mathcal{L_\text{DC}}$ of NIK, we replace the data consistency component to fit a fully-sampled k-space $y$ with an undersampling mask $F$ to the actual acquired k-space $\Bar{y}$ and solve the reconstruction problem: 
\begin{equation}\label{eq:PISCO_fitting}
    \hat{y} = \arg \min_{y} (\lVert \mathrm{F} y - \Bar{y} \rVert_1 + \lambda \cdot \mathcal{ L}_\text{PISCO}(y)).
\end{equation}
% To avoid additional complexity due to density compensation, 
A Cartesian k-space is simulated from a cardiac cine slice (random retrospective undersampling for R=2, 4\% of center lines kept). Optimization is conducted for a total of 500 epochs for two kernel shapes, $[3\times2]$ and $[5\times4]$, with $\mathcal{L}_\text{PISCO}$ weighted by $\lambda$=5e-4. The model is preconditioned for the first 100 epochs and remaining parameters defined as in Sec. \ref{sec:design_choices}.
%%%%%%%%%%%%%%%%%%%%%%%%%% PISCO DESIGN FIGURE %%%%%%%%%%%%%%%%%%%%%%%%%%
\begin{figure*}[!t]
\centerline{\includegraphics[width=\textwidth]{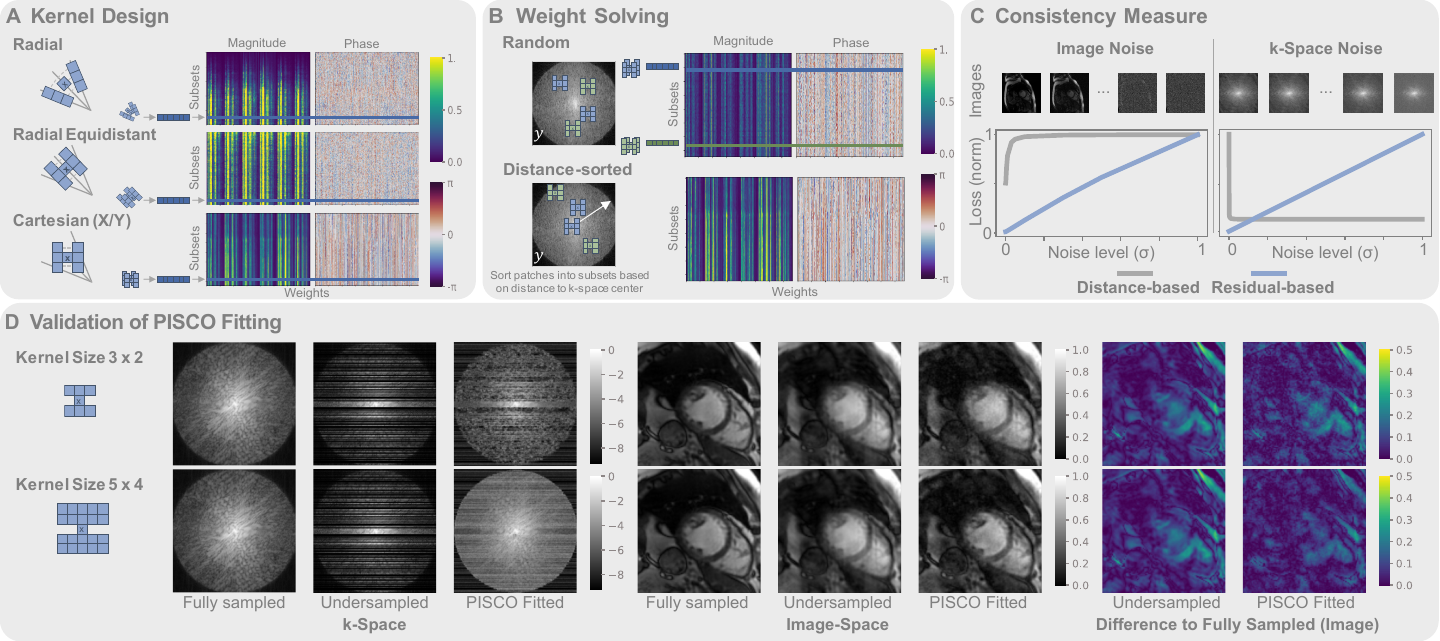}}
\caption{PISCO validation results. 
(A) Kernel Design: For each kernel geometry, multiple subsets of patches (one exemplary shown in blue) are sampled and solved for the weights (dark blue vector). The magnitude and phase of all weight vectors are stacked in the plot. To validate the PISCO condition, all weight vectors should result in the same solution, e.g. a vertical pattern is expected. Only the Cartesian kernel - shown in y-direction, but equal result in x-direction (90° turned) - fulfills this condition.
(B) Weight Solving: Solution of subset weights (magnitude and phase) are stacked vertically (green and blue vectors are two examples). Top: Subsets consist of randomly sampled patches. Bottom: The sampled patches are sorted from the center to the outside of the k-space (white arrow) and then into subsets. This results in subsets with patches of similar k-space magnitude, leading to less noisy weight vector solutions overall.
(C) Consistency Measure: Non-ideal k-spaces are simulated by adding increasing noise in image-space (left) and k-space (right). On the corresponding k-spaces, the distance-based \cite{Spieker_2024_pisco} and residual-based (proposed) PISCO loss is computed and normalized by the maximum loss value. For image noise, both losses increase monotonically, while the residual-based loss is consistently sensitive. For k-space noise, only the residual-based loss monotonically increases, making it feasible for optimization towards the ideal k-space (where $\sigma$=0). 
(D) Validation of PISCO Fitting: K-Space (left) and image (middle) results for PISCO fitted undersampled k-spaces. Within k-space, inclusion of PISCO allows for derivation of missing k-space lines without any additional knowledge. A larger kernel (5x4, bottom) increases the perceptive field and fills more missing k-space lines. Reconstruction results lead to reduced undersampling using PISCO compared to the original undersampled image, visible in the images (middle) as well as difference images (right). 
}
\label{fig:results_piscodesign}
\end{figure*}
%%%%%%%%%%%%%%%%%%%%%%%%%% PISCO DESIGN FIGURE %%%%%%%%%%%%%%%%%%%%%%%%%%

\subsection{Training NIK with PISCO}
\label{sec:training_and_inference}
We adapt NIK's \cite{Huang_2023} architecture using 4 layers, 512 hidden features, high-dynamic range loss as $\mathcal{L_\text{DC}}$, SIREN activations with $\omega$=20 \cite{Sitzmann_2020}, batch size of 10k and use STIFF feature encoding \cite{Catalan_2023} with $\sigma$ as as initialization of the feature distribution. %with spatial percentage of 0.75 for feature encoding. 
Depending on the motion pattern, we rescale the navigator $t$, and correct linear drifts for abdominal imaging (see Table \ref{table:NIK_params}).
%to [0,0.5] or [0,1] for respiratory and cardiac motion, respectively.

For NIK with PISCO, $\mathcal{L}_\text{PISCO}$ is included after $E_\text{pre}$=1000 and weighted by $\lambda$ depending on the dataset and acceleration factor.
A large number of coils $N_c$ within a dataset results in a large number of weight parameters $N_w$. To minimize the amount of patch pairs sampled within each epoch (respective to computational overhead), we adjust the minimum number of sampled subsets $N_{s,\text{min}}$ accordingly. Also, targets for PISCO computation originate from a Cartesian grid ($\delta_{k_{x}}/\delta_{k_y} = N_{FE}^{-1}$) and neighbors are sampled according to kernel design. To account for undersampling in both, x-/y-dimension, the Cartesian kernel design is applied alternating in both directions. 
Weight solving parameters are defined as in Sec. \ref{sec:design_weight}. All models are jointly optimized for a total of 5000 epochs (NVIDIA RTX A6000, Python 3.10.9/PyTorch 1.13.1) with Adam (lr=1e-5), with amsgrad enabled to encounter convergence issues due to the high dynamic range of k-space \cite{Reddi_2019}. 

\renewcommand{\arraystretch}{1.15} % Adjust row spacing
\begin{table}[h]
\caption{Summary of Feature Processing and Loss Parameters}
\centering
\begin{tabular}{cc|C{14mm}|C{16mm}|C{13mm}}   
& & {\textbf{Upper leg}} & \textbf{Cardiac cine}  & \textbf{Abdominal}     \\
\Xhline{1pt} %\hline \hline
\centering
\multirow{3}{*}{\rotatebox[origin=c]{90}{Features}} 
                                        & Drift correction     & - & \ding{55}   & \ding{51}           \\  \cline{2-5}
                                        & Navigator scaling    & -   & {[}0,1{]}          & {[}0,0.5{]}   \\  \cline{2-5}
                                        & $\sigma $             & 6   & 6                  & 1             \\  
\Xhline{1pt} %\hline \hline
\multirow{3}{*}{\rotatebox[origin=c]{90}{PISCO}}    
                                        & $\lambda$               &   0.05  & 0.01-0.15 &   0.01        \\  
                                         \cline{2-5}
                                        & $N_{s, \text{min}}$           & 20  & 30                 & 15            \\
                                         \cline{2-5}
                                        & $f_{OD}$ / $\alpha$     & \multicolumn{3}{c}{1.1 / 1e-4}    \\
                                        % \addlinespace[-2ex]
                                        % \cmidrule(r){3-5} % Line below f_{OD} row, spanning columns 3 to 5
                                        % & $\alpha$             & \multicolumn{3}{c}{1e-4}           \\
                                        % \addlinespace[-2ex]
                                        % \cmidrule(r){3-5} % Line below \alpha row, spanning columns 3 to 5
\Xhline{1pt} %\hline \hline
\end{tabular}
\label{table:NIK_params}
\end{table}

% to match the magnitude of $L_{\text{DC}}$, i.e., $\lambda$=0.01/0.1 for simulation/in-vivo data. 
% For the neighbors P, a kernel of size [3,3] with $\delta x / \delta y = N_{FE}^{-1}$ is sampled around the target T resulting in $N_n$ = 8, as shown in Fig. \ref{fig:overview}. 
% Hence, $N_w$ for the phantom is 288 $(8 \cdot 6 \cdot 6)$ and for the subject would be 5408 $(8 \cdot 26 \cdot 26)$. To increase the number of possible sets for the subject ($N_s \propto N_w^{-1}$), we reduce the number of output coils solved for in each iteration, thereby lowering $N_w$ to 624 $(8 \cdot 26 \cdot 3)$. 
% Due to the large magnitude difference, the center region of k-space (radius=$5\cdot \delta x$) is excluded for weight computation.
% Weight solving for $L_{\text{PISCO}}$ is regularized with $\alpha$=1e-4 and $f_{OD}$=1.1. 
% Both losses are optimized with separate Adam optimizers (lr=3e-5). All models are run for 
% 1000 epochs total (NVIDIA RTX A6000, using Python 3.10.1/PyTorch 1.13.1).

\subsection{Baseline Comparisons and Evaluation Metrics}
\label{sec:eval}
To evaluate the performance of our proposed PISCO regularization, we compare standard NIK \cite{Huang_2023}, PISCO-dist \cite{Spieker_2024_pisco} and PISCO (NIK with $\mathcal{L}_\text{PISCO-dist}$ and $\mathcal{L}_\text{PISCO}$, respectively). Further, we compute the inverse NUFFT (INUFFT) as well as the state-of-the-art motion-resolved reconstruction method XD-GRASP \cite{Feng_2016}. For the latter two, the number of motion states (MS) is defined depending on the dataset, i.e., 1MS for static upper leg, 25MS for cardiac cine, and 4/50MS for abdominal (in the following referrered to as INUFFT\textit{MS} or XD-GRASP\textit{MS}). We conduct the XD-GRASP reconstruction using a conjugate gradient algorithm with line search and, for each dataset, perform a grid search on a representative subject to determine the TV regularization weight $\lambda_{TV}$ (see Table \ref{table:imaging_parameters}). 

% \subsection{Evaluation}
% \label{sec:eval}
Additionally, quantitative metrics such as peak signal-to-noise ratio (PSNR) and spatial feature similarity (FSIM-spat) of all time points are evaluated. The dynamic performance is compared using the temporal feature similarity (FSIM-temp) on all $xt/yt$ slices. Before metric computation, images are clipped to their $99^{th}$ percentile and normalized to [0,1]. 

%%%%%%%%%%%%%%%%%%%%%%%%%%%%%%%%%%%%%%%%%%%%%%%%%%%%%%%%%%%%%
\section{Results}

\subsection{PISCO Design Choices}
\label{sec:results_piscodesign}
\textbf{Kernel Design} Fig. \ref{fig:results_piscodesign}A shows the visualization of the weight vector solutions $ \mathrm{W}_s$ (individual weights on the horizontal axes) for multiple randomly sampled subsets $s=1...N_s$ (stacked on vertical axes) for all three kernel designs. Since the weights were computed on an ideal k-space, all subset weight vectors should be the same according to PISCO, i.e., a vertical pattern should be visible. No consistent vertical pattern is visible for the radial as well as radial equidistant kernel. Yet, only the Cartesian kernel results in consistent magnitudes and phases for all subsets, confirming the applicability of PISCO as self-supervised consistency measure with this kernel design. 

% Weight Solving
% \subsubsection{Weight Solving}
\textbf{Weight Solving} The subset weight vector solutions with and without distance sorting from the center to periphery k-space are shown in Fig. \ref{fig:results_piscodesign}B. Both weight vector solutions indicate that including Tikhonov regularization and the overdetermination factor result in stable and consistent solutions. Yet, more noise in both, the magnitude and phase of the weight vector solution, is recognizable when no frequency sorting is applied. Avoiding poorly scaled LES by sorting the patch pairs according to their k-space distance results in less variance of the $\mathrm{W}_s$, and hence, a more stable PISCO condition.

% Consistency Measure
% \subsubsection{Consistency Measure}
\textbf{Consistency Measure} As shown in Fig. \ref{fig:results_piscodesign}C, both PISCO measures (distance and residual-based) increase monotonically with rising noise level in image-space. Yet, the linear increase of residual PISCO may offer additional stability when optimizing with the objective of a noise-free solution compared to the distance-based metric, where the gradient is sensitive to the amount of noise added. For k-space added noise, the distance-based measure decreases with increasing k-space noise, which makes this measure infeasible for k-space noise reduction optimization. In contrast, a proportional relationship of residual PISCO to k-space based noise can be observed, making it a suitable metric for further optimization. TO further validate this finding, we include reconstructions using the distance-based loss (PISCO-dist) in the in-vivo results.

%%%%%%%%%%%%%%%%%%%%% PISCO Convergence %%%%%%%%%%%%%%%%%%%%%%%%%%%%%%%%%%%
\subsection{Validation of PISCO Convergence}
The undersampled k-space fitting results using PISCO with two different kernel sizes are shown in Fig. \ref{fig:results_piscodesign}D. Without any additional information, PISCO is capable of filling the undersampled gaps within k-space (Fig. \ref{fig:results_piscodesign}D left). The larger kernel enables a larger receptive field and consequently, less remaining gaps within k-space. The corresponding reconstructions show sharper results and higher quantitative values when including PISCO (Fig. \ref{fig:results_piscodesign}D middle) as well as reduced undersampling artefacts, as visible in the difference images (Fig. \ref{fig:results_piscodesign}D right). Note that this reconstruction does not leverage any temporal redundancy yet, but the improvement is solely based on including the neighborhood relationship with PISCO.

%%%%%%%%%%%%%%%%%%%%% PISCO for NIK %%%%%%%%%%%%%%%%%%%%%%%%%%%%%%%%%%%
\subsection{PISCO for NIK Regularization}
\begin{figure}[h]
\centerline{\includegraphics[width=\columnwidth]{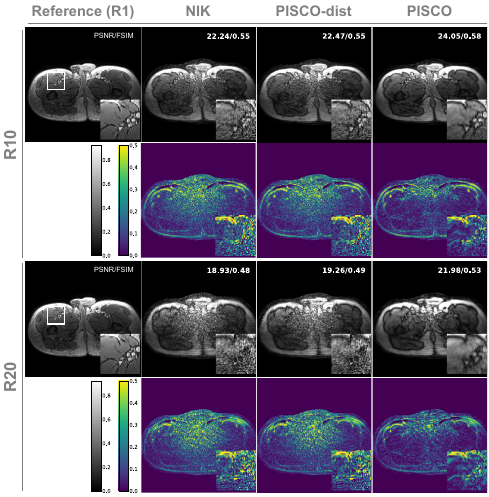}}
\caption{Static upper leg: Qualitative and quantitative results for two exemplary acceleration factors (R10/20). NIK \cite{Huang_2023} results in increasing noise, particularly towards the center of the reconstruction. PISCO-dist \cite{Spieker_2024_pisco} marginally improves reconstructions, but still remains noisy. With the proposed PISCO a better implicit representation is learned, that results in reduced noise 
%(PSNR~\texttt{$\uparrow\!$}) 
and sharper structures (PSNR and FSIM~\texttt{$\uparrow\!$}). 
% Column width in cm: \printlength{\columnwidth}~cm
}
\label{fig:results_hip}
\end{figure}

%%%%%%%%%%%%%%% Knee %%%%%%%%%%%%%%%
\textbf{Upper leg} Quantitative and qualitative results are visualized in Fig. \ref{fig:results_hip} for two acceleration factors. NIK's reconstruction performance noticeably degrades with increasing acceleration, i.e. reduced amount of training data. PISCO-dist marginally improves reconstructions and metrics, but remains noisy. In contrast, the proposed PISCO results in less noisy reconstructions than both, NIK and PISCO-dist. Also, it recovers vessel details more reliably, represented in higher PSNR and FSIM, respectively. Again, no temporal redundancy is leveraged yet.
% Comparison NUFFT vs. NIK vs. PISCO - R5/10/20 -> improvement

\begin{figure*}[!t]
\centerline{\includegraphics[width=\textwidth]{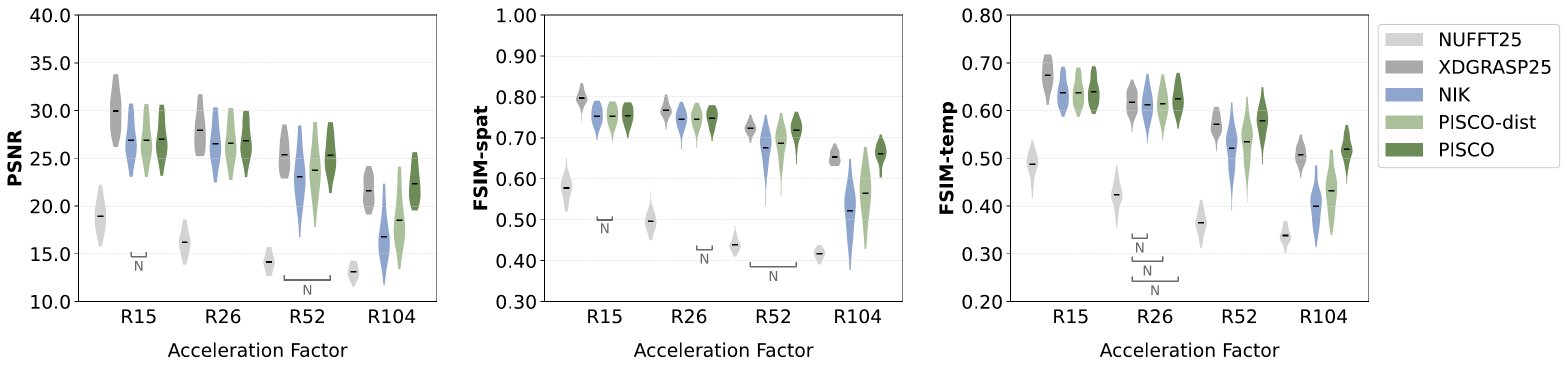}}
\caption{Cardiac cine: Quantitative reconstruction results %(PSNR/FSIM-spat/FSIM-temp) 
of 30 subjects for acceleration factors R15/R26/R52/R104 (with $\lambda$=0.01/0.05/0.1/0.15 for PISCO-dist/PISCO reconstruction, respectively). For R15, XD-GRASP25 offers least noisy reconstruction (PSNR) with high spatial and temporal resolution (FSIM-spat/FSIM-temp). For R26, NIK, PISCO-dist and PISCO lead to similar temporal results. At even larger acceleration factors, XD-GRASP25 performance decreases, as does NIK rapidly. PISCO-dist outperforms NIK, but does not reach XD-GRASP25. Yet, inclusion of the proposed PISCO significantly improves temporal structure compared to all other methods (R$>$26). At the same time, spatial reconstruction quality is maintained (R52) or even significantly improved (R104) compared to XD-GRASP25. 
All comparisons, except those marked with "N", are statistically significant (Wilcoxon signed rank test with False Discovery Rate correction at p$<$0.05). }
\label{fig:results_cardiac_quant}
\end{figure*}

%%%%%%%%%%%%%%% Cardiac %%%%%%%%%%%%%%%
\textbf{Cardiac Cine}
% Quantitative
Quantitative results of the cardiac cine dataset (Fig. \ref{fig:results_cardiac_quant}) show that PISCO consistently outperforms NIK and PISCO-dist at all acceleration factors in both, spatial and temporal metrics. Particularly at high acceleration factors (R52/R104 or 4/2 spokes per frame), NIK's and PISCO-dist's spatial and temporal performance drastically decay. In contrast, PISCO enables spatial reconstruction quality similar or better to the reference method XD-GRASP25 (PSNR/FSIM-spat) and additionally, models the temporal dynamics better at these high accelerations (FSIM-temp). 

\begin{figure*}[h]
\centerline{\includegraphics[width=\textwidth]{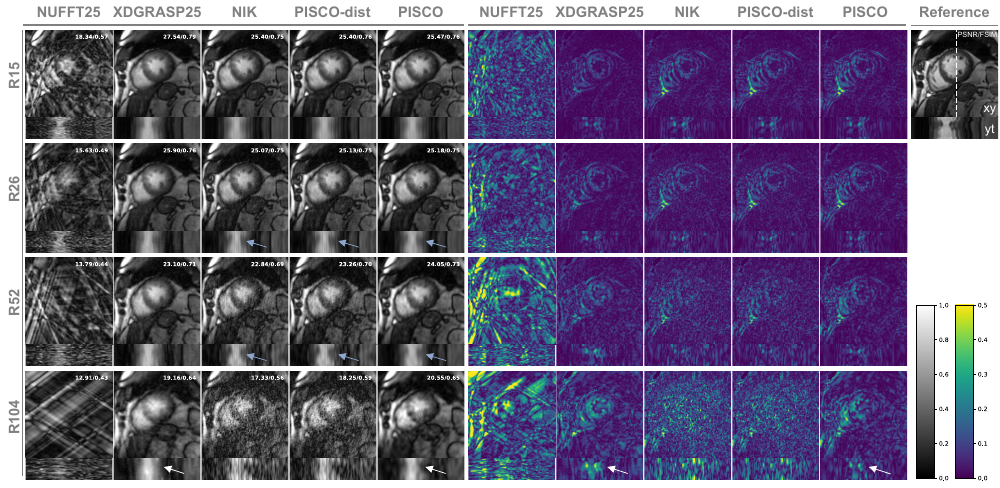}}
\caption{Cardiac cine: Qualitative reconstruction results of one subject for four acceleration factors. The reference reconstruction (R1) is shown in the top right (xy), with the white dotted line marking the slice where the temporal profile (yt) is extracted. Videos of the complete dynamic reconstruction (xyt) can be found in the Supplementary (Figure 1). 
Similar to the quantitative results, XD-GRASP25 offers high spatial resolution by applying smoothing in the temporal domain, resulting in decreasing temporal reconstruction quality with increasing accelerations (i.e. R52/104, white arrow). When acceleration increases, NIK \cite{Huang_2023} and PISCO-dist \cite{Spieker_2024_pisco} suffer from increasing noise in the spatial and temporal dimension (see difference images R52/R104 and blue arrows, respectively). Particularly for high accelerations (R52+), the proposed PISCO results in similar spatial and improved temporal quality compared to XD-GRASP25 (white arrows). 
}
\label{fig:results_cardiac_qual}
\end{figure*}

% Qualitative
Similar observations can be made in the qualitative reconstructions of one exemplary subject (Fig. \ref{fig:results_cardiac_qual}). All motion-resolved reconstruction methods, XD-GRASP25, NIK, PISCO-dist and PISCO, encounter the strong undersampling artefacts visible in INUFFT. XD-GRASP results in spatial smoothness by regularizing over the temporal dimension, which introduces blurring, particularly with reduced data (R52/R104). NIK and PISCO-dist result in noisy spatiotemporal reconstructions, particularly with increasing acceleration factors. With PISCO, an improved neural k-space representation could be learned, that is spatially smooth, and recovers temporal dynamics even at 2 spokes/frame (R104). At high acceleration rates (R52/R104), PISCO surpasses the state-of-the-art XD-GRASP25 in capturing temporal detail (FSIM-temp) while achieving comparable spatial smoothness (PSNR/FSIM-spat). Note that in this case, only 25 time points were analyzed due to the use of binned reference data. Yet, PISCO's temporal resolution can further be increased by sampling more temporal points. Nonetheless, at accelerations like R104, the resulting images do not yet reach diagnostic quality but may be valuable for intermediate applications, such as motion estimation.

%%%%%%%%%%%%%%% Abdominal %%%%%%%%%%%%%%%
\textbf{Abdominal}
\begin{figure*}[!t]
\centerline{\includegraphics[width=\textwidth]{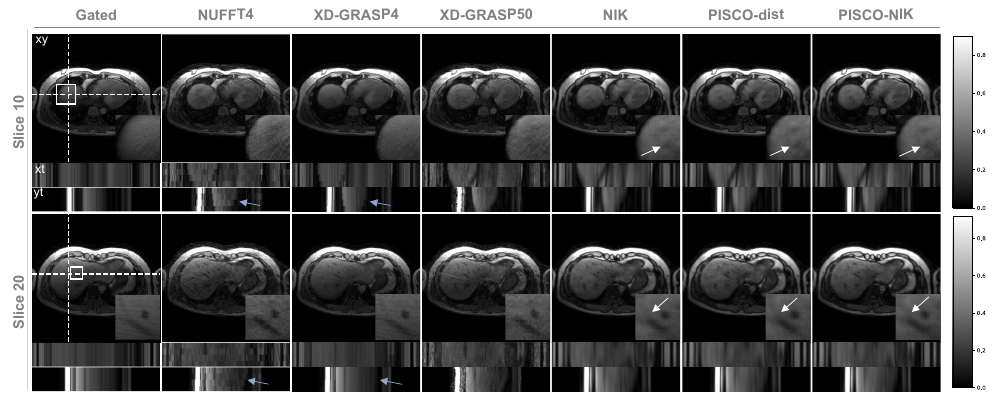}}
\caption{Abdomen - respiratory resolved: Qualitative reconstruction results for two slices of a subject are shown: the xy-image depicts spatial reconstruction, while temporal profiles (marked by a white dotted line) indicate temporal reconstructions (xt/yt) below. Videos of the complete dynamic reconstruction (xyt) can be found in the Supplementary (Figure 2). 
The gated reconstruction (unpaired to the rest) does not offer any temporal information and remains blurry (zoom-in). Although XD-GRASP4 improves spatial resolution compared to NUFFT4, its temporal resolution remains limited (blue arrows). Increasing the number of MS (XD-GRASP50) results in better temporal resolution, but noisier spatiotemporal reconstructions. NIK reduces the noise in the temporal dimension. PISCO-dist and PISCO further smoothen spatiotemporal reconstructions, while the proposed PISCO maintains clearer vessel structures (white arrows).}
\label{fig:results_abdominal}
\end{figure*}
Fig. \ref{fig:results_abdominal} shows reconstructions of exemplary slices of the abdominal data. The reference unpaired prospectively GATED acquisition, common in a clinical setting, shows a spatial smooth image, but still appears slightly blurry at the organ edges and lacks temporal information overall. Motion-resolved reconstructions with 4MS retain undersampling artefacts (INUFFT4) or lose temporal information (XD-GRASP4). Increasing the temporal resolution by binning to 50 MS (XD-GRASP50) improves the dynamic depiction, but suffers from noise and undersampling. NIK, PISCO-dist and PISCO achieve high spatiotemporal resolution, with PISCO further smoothing results both spatially and temporally.

%%%%%%%%%%%%%%%%%%%%%%%%%%%%%%%%%%%%%%%%%%%%%%%%%%%%%%%%%%%%%
\section{Discussion}

%%%%%%% Short Summary %%%%%%%
Based on the concept of parallel imaging-inspired self-consistency (PISCO), we have proposed a novel k-space consistency measure which can be determined in a self-supervised manner. With multiple ablation/simulation studies, we have validated the convergence behaviour of PISCO, and hence, its applicability as objective function within dynamic MR reconstruction, i.e. using NIK. Without the need for any additional data, we have verified PISCO's potential to learn improved neural implicit representation, resulting in enhanced spatial and temporal image quality.

%%%%%%% Observations on current data %%%%%%%
%% Knee
\paragraph{PISCO for Improved MR Reconstruction}
While NIK was originally developed for dynamic reconstruction, PISCO is not limited to this application. For both, simple k-space fitting as well as learning a static NIK, PISCO enables reduction of undersampling artefacts. Still, performance improvement will always be limited since no additional information except the neighborhood constrain is available and any type of redundancy, i.e. given by the temporal dimension or multiple echoes \cite{Spieker_2024_denik}, is expected to further improve PISCO's positive impact as well.
% NIK generally profits from temporal redundancy, undersampling in staic example -> no additional information, PISCO improves NIK, but not necessarily better than NUFFT (cause no source of information on how to interpolate regions)

% Cardiac
Incorporating the temporal dimension enhances representation learning for higher acceleration factors (e.g., achieving R$\geq$26 in cardiac cine imaging versus R20 in upper leg imaging). Yet, NIK also struggles when very little information is available per frame (R$\geq$52, or $\leq$4 spokes per frame), likely due to overfitting to noise in unsampled k-space regions. The proposed PISCO adresses this overfitting noise by (1) including the unknown points in the training procedure and (2) enforcing consistency throughout all the k-space points, visible in k-space and image-space. For lower acceleration factors, XD-GRASP remains a viable comparison but tends to oversmooth at R$\geq$52 (Fig. \ref{fig:results_cardiac_qual}), sacrificing temporal information. In contrast, PISCO-NIK maintains temporal resolution without such trade-offs. 
Still, PISCO's performance remains limited by the binned, discontinuous cardiac cine dataset, which inherently introduces residual motion blurring during training. 
Future studies on real time cardiac data
% , which avoid the inherent residual motion blurring during training with the given binned dataset, 
are anticipated to further demonstrate PISCO-NIK’s improved efficiency.

%% Abdominal
In abdominal imaging, although continuous data is available, the irregularity of respiratory motion compared to cardiac motion presents additional challenges. Acquired spokes are unequally distributed between end-exhale and end-inhale phases. Also, higher resolution and expanded field of view (FOV) requirements increase k-space gaps in radial acquisitions, complicating reconstruction. This presents visible challenges for XD-GRASP4 (Fig. \ref{fig:results_abdominal}), which prioritizes spatial resolution in abdominal imaging but sacrifices dynamic information, further highlighting the potential advantages of PISCO-NIK in this application. Yet, PISCO-NIK's performance is highly dependent on a reliable navigator signal, and additional research is needed to enhance its robustness and improve hysteresis effect modeling to refine respiratory-resolved reconstructions.

%%%%%%% Benefits and further potential
\paragraph{General PISCO Design}
% Ablations
Within our PISCO validation studies (Sec. \ref{sec:results_piscodesign}), we have shown a feasible kernel, weight solving and consistency measure design to leverage the global neighborhood relationship for k-space optimization. While a Cartesian kernel design has proven crucial for global consistency (Fig. \ref{fig:results_piscodesign}A), flexibility remains regarding kernel size and distance. In the k-space fitting example (Fig. \ref{fig:results_piscodesign}D), a larger kernel resulted in even better reconstruction results. Yet, an increased kernel size also results in a larger computational overhead, since the number of weights $N_w$ proportionally increases, and hence, the required number of patches to solve one subset. A potential solution could be a more efficient handling of the coil dimension, since $N_w \propto N_c^2$. A computationally efficient solution for weight solving would also open possibilities for advanced physical modeling with PISCO in the temporal dimension, such as motion, time-dependent MR field imperfections \cite{Wang_2019,Abraham_2023} or phase modeling \cite{Haldar_2016}.
% what about adding t as kernel dimension, \cite{Wang_2019} does this for estimation of B0 effects - would increase the amount of params needed and require the same relative relationship between consecutive motion states?

%%%%%%% PISCO-dist vs PISCO
Regarding the consistency measure, we have validated the improved convergence behaviour of the residual-based PISCO measure vs. the previously proposed distance-based loss (PISCO-dist in reconstruction results). The reduced sensitivity to image noise (Fig. \ref{fig:results_piscodesign}C) is also evident in the in-vivo reconstructions (Fig. \ref{fig:results_hip}/Fig. \ref{fig:results_cardiac_qual}), where PISCO-dist offers only marginal denoising improvements, likely due to suboptimal optimization. An additional benefit of the proposed PISCO measure is the reduced computational effort, since expensive distance calculations between all weight vectors (with thousands of complex numbers) are avoided.

% Fitting to individual dataset
While some parameters, such as those for weight solving, were generally applicable across datasets, others require adaptation to specific applications. One example is the weighting factor for the PISCO loss, $\lambda$, which has been adapted by observing PISCO's loss magnitude and behaviour during optimization. Extending the proposed residual-based loss to automatically adapt to the dataset - considering factors such as kernel design, $N_c$, and acceleration - could further enhance PISCO's out-of-the-box applicability.
% Some parameters need to be fitted to dataset, automatic determination with further analysis of loss properties, e.g. weighting by number of samples etc

The observed NIK reconstruction improvements at high acceleration factors suggest potential of PISCO for reducing clinical protocol times. Its efficient design also allows flexible integration with other regularization methods, offering opportunities for further improvements in reconstruction quality. Moreover, it may open new avenues for rapid motion estimation, which could be beneficial for subsequent analysis and downstream tasks.
%%  While the images shown are not necessarily of diagnostic quality, the considerable improvement suggests PISCO's potential for spatiotemporal k-space regularization. Can be combined with further k-space constraints such as LORAKS \cite{Haldar_2016}. 

%% Complex-valued design beneficial for reconstructions?

%%%%%%% General Limitations
% Stopping criterion, when to stop training ->> the longer the smaller the loss but also risk of overfitting
% -> PISCO already helps with increasing training speed, but stopping criterion unknown
% Uncertainty and training times

%%%%%%%%%%%%%%%%%%%%%%%%%%%%%%%%%%%%%%%%%%%%%%%%%%%%%%%%%%%%%
\section{Conclusion}
With PISCO, we have demonstrated how a conventional parallel imaging concept can be adapted into a self-supervised consistency measure that enhances learning-based MR reconstruction. Its calibration-free and flexible design allows for seamless integration into the training process, making it a promising method for application in other anatomies or k-space based reconstruction techniques. 
% The observed NIK reconstruction improvements at high acceleration factors suggest potential for reducing clinical protocol times and open new avenues for rapid motion estimation, which could be beneficial for subsequent analysis and downstream tasks.
% A conclusion section is not required. Although a conclusion may review the main points of the paper, do not replicate the abstract as the conclusion. A conclusion might elaborate on the importance of the work or suggest applications and extensions.

%%%%%%%%%%%%%%%%%%%%%%%%%%%%%%%%%%%%%%%%%%%%%%%%%%%%%%%%%%%%%
\section*{Acknowledgment}
V.S. and H.E. are partially supported by the Helmholtz Association under the joint research school ”Munich School for Data Science - MUDS”. V.S. conducted part of the research during a visit at iHEALTH as Bayer Foundation Fellow.

\printbibliography

\end{document}